\documentclass[%
 reprint,
showkeys,
 amsmath,amssymb,
 aps,
]{revtex4-2}

\usepackage{graphicx}
\usepackage{dcolumn}
\usepackage{bm}
\usepackage{siunitx}
\usepackage{amsfonts,amssymb}
\usepackage[colorlinks=true, allcolors=blue]{hyperref}
\usepackage{graphics}
\usepackage{rotating}
\usepackage{amsmath}
\usepackage{subcaption}
\usepackage{xcolor}

\begin{document}

\preprint{APS/123-QED}

\title{Synthesis of cold and trappable fully stripped HCI’s via antiproton-induced nuclear fragmentation in traps}

\author{G. Kornakov$^{1*}$}
\author{G. Cerchiari$^{2}$}
\author{J. Zieli\'nski$^{1}$}
\author{L. Lappo$^{1}$}
\author{G. Sadowski$^{3}$}
\author{M. Doser$^{3}$}

\affiliation{$^{1}$Warsaw University of Technology, Faculty of Physics, ul. Koszykowa 75, 00-662 Warszawa, Poland}
\affiliation{$^{2}$Institut f\"ur Experimentalphysik, Universit\"at Innsbruck, Technikerstrasse~25, 6020~Innsbruck, Austria} 
\affiliation{$^{3}$CERN, Esplanade des particules 1, 1211 Geneva, Switzerland}
\affiliation{$^{*}$Contact address: georgy.kornakov@pw.edu.pl}





\begin{abstract}
The study of radioisotopes as well as of highly charged ions is a very active and dynamic field. In both cases, the most sensitive probes involve species trapped in Penning or Paul traps after a lengthy series of production and separation steps that limit the types and lifetimes of species that can be investigated. We propose a novel production scheme that forms fully (or almost fully) stripped radionuclei in form of highly charged ions (HCI's) directly in the trapping environment. The method extends the range of species, among them radioisotopes such as $^{21}$F, $^{100}$Sn or $^{229}$Th, that can be readily produced and investigated and is complementary to existing techniques.
\end{abstract}

\maketitle


\section{Introduction}

The formation of radioisotopes currently relies on proton-induced spallation (isotope separators), fragmentation (FRIB)~\cite{FRIB} or naturally occurring decay chains. In the case of highly charged ions (HCI's), their formation requires stripping (either through collisional processes in a series of foils or through interaction with an intense current of high energy electrons, in which the majority or totality of electrons can be removed from a beam of accelerated singly ionized atoms), before the resulting HCI's are trapped and cooled for subsequent study. In all cases, the involved formation processes set constraints on the types of species that can be produced or on the lifetimes of the isotopes to be investigated. In order to either provide radioisotopes that are currently difficult to produce or extract and trap, or HCI's of short-lived radioisotopes (in contrast to stable or very long-lived isotopes), new approaches are needed. 

In this paper, we extend and investigate a scheme proposed in Ref.~\cite{Gerber} for pulsed formation of protonium, an exotic hydrogen-like atom consisting of a proton and an antiproton instead of an electron, to a wide range of possible alternative precursor elements, ranging from light to very heavy atoms, and follow the subsequently produced antiprotonic atoms through the atomic cascade, annihilation on the surface and possible fragmentation stages to the fate of any produced nuclear remnants. This exploration is carried out via the GEANT4~\cite{GEANT4:2002zbu,Allison:2006ve,Allison:2016lfl} simulation package for several underlying physics descriptions of the annihilation and fragmentation stages, while relying on established calculations for atomic formation and cascade.

We first sketch the physics processes involved in this proposed production scheme in Sections~\ref{sec:Physics_atomic} and \ref{sec:Physics_nuclear}, before focusing on the simulation of the annihilation and fragmentation process. For a selection of precursor elements (F, Co, Ho, Ta, W, Os, Au, Po, Rn, Ra, Th, U, Pu), 
the yields, trappability and characteristics of the produced nuclear fragments are explored and compared to existing production protocols in Section~\ref{Results}. 
The systematic uncertainty on the expected relative yields of trappable, fully stripped radioisotopes, 
as well as of the implications of the technique are addressed in Section~\ref{Discussion}. 
Conclusions are drawn in Section~\ref{Conclusions} where we also suggest the next steps required in validating the production process proposed here for trapped, fully stripped HCI's of (also short-lived) radioisotopes and establishing the viability of the approach. 

\section{Formation of antiprotonic atoms}\label{sec:Physics_atomic}

The formation of antiprotonic atoms has in the past involved injecting low-energy antiprotons into gaseous, liquid, or solid target materials~\cite{Widmann1995quenching}, in which antiprotons undergo further energy loss through collisional interactions with the target material's atomically-bound electrons. 
Once the velocity of the antiprotons matches that of a less bound electron, the antiproton is captured and replaces this electron. 
This process results in the formation of an antiprotonic atom with the antiproton being in a highly excited Rydberg state \cite{Cohen} due to the large difference in mass between the two negatively charged particles. 

The $\overline{\textrm{p}}$A* lifetime is determined by the relaxation process that brings the antiproton's wavefunction to have a substantial overlap with the nuclei, resulting in the annihilation of the antiproton with one of the surface nucleons~\cite{Backenstoss_1989}.
The cascade of the $\overline{\textrm{p}}$ inside the atom follows two different regimes depending on the principal quantum number $n$ and the angular momentum quantum number $l$ of the initial orbital. 
For example, in $\overline{\textrm{p}}$--He it has been shown that for $n,l\gtrsim38$ the Rydberg state decays via radiative de-excitation of the orbital that can last a few \SI{}{\micro\second}~\cite{Iwasaki_1991_PRL_discovery_antiprotonic_helium}. 
However, combining other molecules to He, these states are also susceptible to rapid quenching induced by collisions of the antiprotonic atoms in the target medium of formation~\cite{Yamazaki1993quenching}. 
For lower quantum numbers of the orbital state, the $\overline{\textrm{p}}$ quickly cascades inside the atom, Auger-ejecting the electrons on its way to the nucleus within a few hundreds of \si{\ps} or annihilates from high-nS states~\cite{Backenstoss_1989}. 
For electron binding energies of less than about 40 keV, e.g. in $\overline{\textrm{p}}$-Kr, stripping is complete; for larger binding energies, e.g. in $\overline{\textrm{p}}$-Xe, a small number of deeply bound electrons may 
remain~\cite{PhysRevA.38.4395, EPJD.47.11}.
When these processes occur in bulk matter, the annihilation products follow a trajectory from their formation point until they reach either the container's surface from which they can emerge or they are absorbed within the medium of the same bulk matter and cannot be studied, limiting the knowledge of the spectra of produced heavy fragments. 

\begin{figure}[t]
\centering
\includegraphics[width=0.48\textwidth]{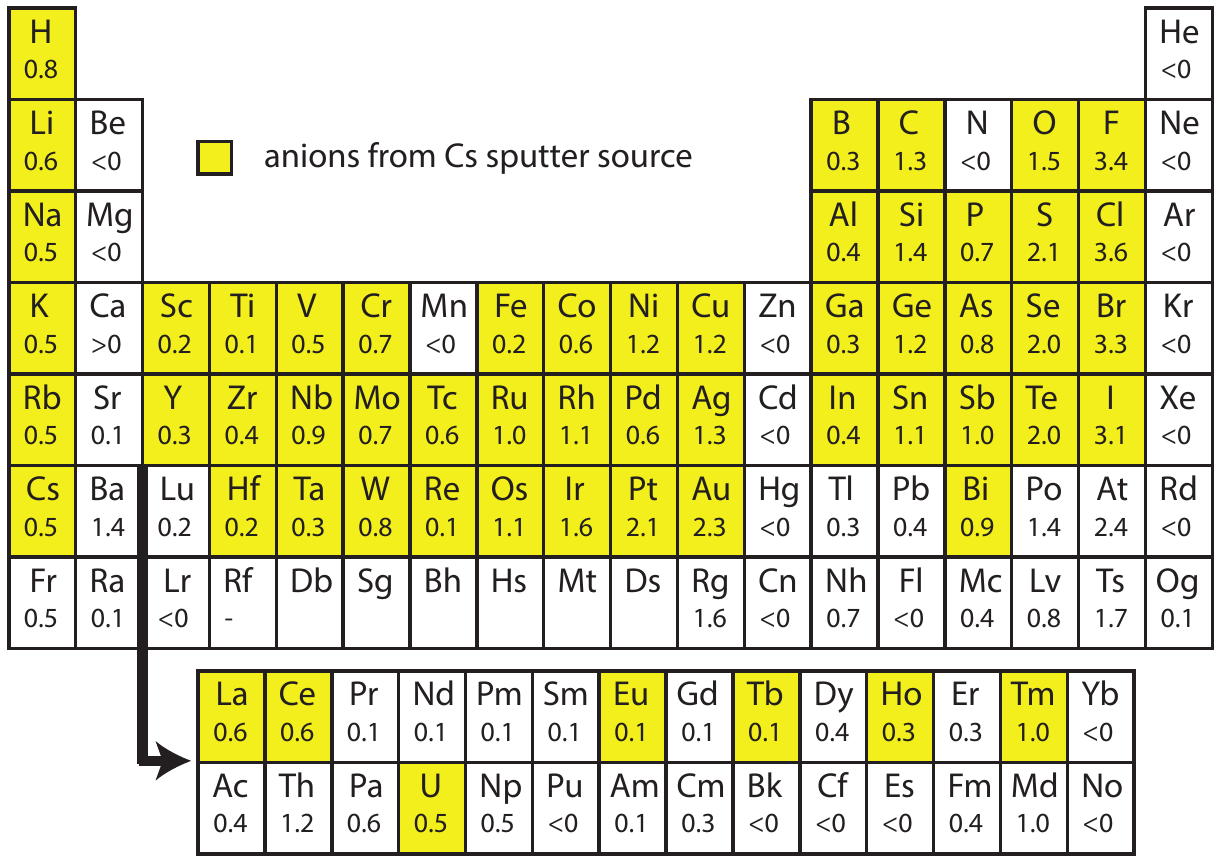}
\caption{Periodic table with the electron affinity of each element in eV. The elements highlighted in yellow can be produced as an elemental anion current from a Cs sputter source~\cite{Middleton_cookbook}.}
\label{fig:anionstable}
\end{figure}

Trapping of massive annihilation products becomes feasible if the formation of antiprotonic atoms takes place in ultra-high vacuum (UHV) and within Penning traps, requiring that both the antiprotons and the atomic precursors interact at low momentum in the trap~\cite{Ordonez2008}. 
This could be realized in a cold antimatter experiment by combining positive or negative ions with antiprotons. 
Antiprotons can be mixed with cations in a nested Penning trap, leading to the formation of antiprotonic atoms via three-body interactions as demonstrated with antihydrogen~\cite{Amoretti2002}, having however a low production cross section. 

Alternatively, anions can be used for producing antiprotonic atoms either by streaming an anion beam on a target of trapped antiprotons or by co-trapping anions along with antiprotons~\cite{Gerber}. 
The anions can be produced outside the trap in form of a beam and subsequently be guided on-axis into the trap as demonstrated in Ref.~\cite{Cerchiari2018}. 
This approach is particularly promising for three reasons. 
First, it allows selecting the atomic element to be interacted among many candidates across the period table. 
If a Cs sputter source~\cite{Middleton_cookbook} is used to form the beam, the elements highlighted in Fig.~\ref{fig:anionstable} become available: the different species are accessible just by substituting the sputtered material in the source's target. 
Second, the UHV condition is preserved by guiding the anions from outside the trap instead of producing them in-situ~\cite{Li1997}. This prevents the antiprotons from annihilating with the rest gas and allows one to separate the antiprotons and the highly-charged nuclear fragments that are subsequently produced in different trapping regions. It furthermore ensures that electron replenishing of these nuclear fragments via collisions with rest gas is minimized, thus extending their lifetime.
Third, each annihilation succeeding the synthesis of an antiprotonic atom can be time tagged by detecting with high efficiency the coincident emission of multiple pions.

In co-trapping experiments involving anions, producing cold antiprotonic atoms implies sympathetic cooling and subsequent photo-detachment of the negative ions~\cite{Gerber}. 
Sympathetic cooling of anions and antiprotons is efficiently achieved by mixing them with an electron plasma~\cite{Andresen2011} to reduce the thermal excitation of antiprotons and atomic anions to between 1 and 100~K. 
This can be realized because anions, electrons and antiprotons have the same electric charge that prevents annihilation by Coulomb repulsion at low energy while allowing their simultaneous manipulation~\cite{Kellerbauer_2006}. 
After the preparation, laser excitation is used to photo-detach the extra-electron from anions and excite the neutral atoms to Rydberg levels with a two-level scheme. 
Photo-detachment removes the Coulomb barrier and the subsequent Rydberg excitation of the neutral atoms favors the resonant charge-exchange interaction between neutralized cold atoms and co-trapped antiprotons. 
The process forms antiprotonic atoms as in bulk but in UHV and at the cryogenic temperatures of the trap environment. Also, in these conditions, the formation cross section can be increased by 6 orders of magnitude with respect to formation in bulk~\cite{Krasnicky2016}.   

In appendix A, we discuss how the production of antiprotonic atoms can be realized by streaming an anion beam on trapped antiprotons. The production scheme involving co-trapping and Rydberg excitation is already discussed in detail in Ref.~\cite{Gerber} for the case of protonium and can be easily extended to any other negative ions.



\section{Nuclear processes within antiprotonic atoms}\label{sec:Physics_nuclear}

Independent of the specific formation process however, at the end of the atomic de-excitation cascade, annihilation between the antiproton and a nucleon of the atom's nucleus takes place in an overall highly charged system. 
Both data from experiments carried out at LEAR~\cite{lubinski, Lubinski94, Lubinski98}  and GEANT4~\cite{GEANT4:2002zbu} simulations indicate that at the end of the atomic cascade of antiprotonic atoms, this annihilation occurs at the periphery of the nucleus, resulting in little recoil momentum.  

Charged pions (or very rarely, kaons) produced during this nucleon-antiproton annihilation~\cite{Polster:1995up} may interact with the remnant nucleus, and result in emission of small numbers of protons, neutrons or alphas, in addition to possibly fragmenting the remnant; a range of low energy, potentially trappable radioisotopic HCI's can thus be produced by evaporation of the heated nuclei caused by the annihilation~\cite{Goldenbaum:1996ih}. 
While only the long-lived (radioactive) remnants have been amenable to identification (through radiochemistry~\cite{Lubinski94}, as experiments to date have only irradiated bulk matter with antiprotons), the resulting elemental distributions are in broad agreement with calculations and simulations of the annihilation and fragmentation process that incorporates both interaction cross sections of mesons with nuclei as well as nuclear de-excitation models~\cite{Aghion:2017xor}. 

To date, with very few exceptions~\cite{MARKIEL1988445, Aghion:2017xor}, the energy distribution of the resulting remnants has not been amenable to measurement; as the simulations developed for this article show, these would not have had sufficient energy to leave the bulk matter environment in which they have been formed, and would have travelled only a few microns in it. The only possibility to perform a complete measurement of the slow remnants is to perform the experiment in conditions of UHV, avoiding any absorption by the environment.

\section{Simulation results} \label{Results}

Radioisotope yields have been determined for a number of stable (F, Co, Ho, Ta, W, Os, Au) or nearly stable (Po, Rn, Ra, Th, U, Pu) target isotopes via a dedicated GEANT4 simulation that assumes that antiprotons have been captured by the target atom. 
In practice, antiprotons are shot at nm-thin solid targets of the specific element at very low energy ($\sim$ 1~keV), in which they are slowed down via standard electromagnetic interactions. 
Once they are at rest, both the annihilation of the antiproton with a nucleon and the nuclear processes that are described by using the GEANT4  FTFP\_BERT\_HP physics list  \cite{antimatter_geant4} that assumes a quark gluon string model for high energy interactions of protons, neutrons, pions and kaons with nuclei, are simulated. 
The excited nucleus that is created via such high energy interactions is passed to a so-called FTFP (FriTioF plus Precompound) model that models the nuclear de-excitation.
Then, for each capture and annihilation event, relevant information as the particle number, three-momentum, energy, mass and charge of the final-state products are stored for subsequent analyses.
Currently, there are no alternative models for interaction at rest of negative hadrons with nuclei as the previously available CHIPS~\cite{Degtyarenko:2000ks}  was abandoned already in GEANT4 version 10.0 (2013).  

\begin{figure}
\centering
\includegraphics[width=0.47\textwidth, trim={5.5cm 0 7.5cm 0}]{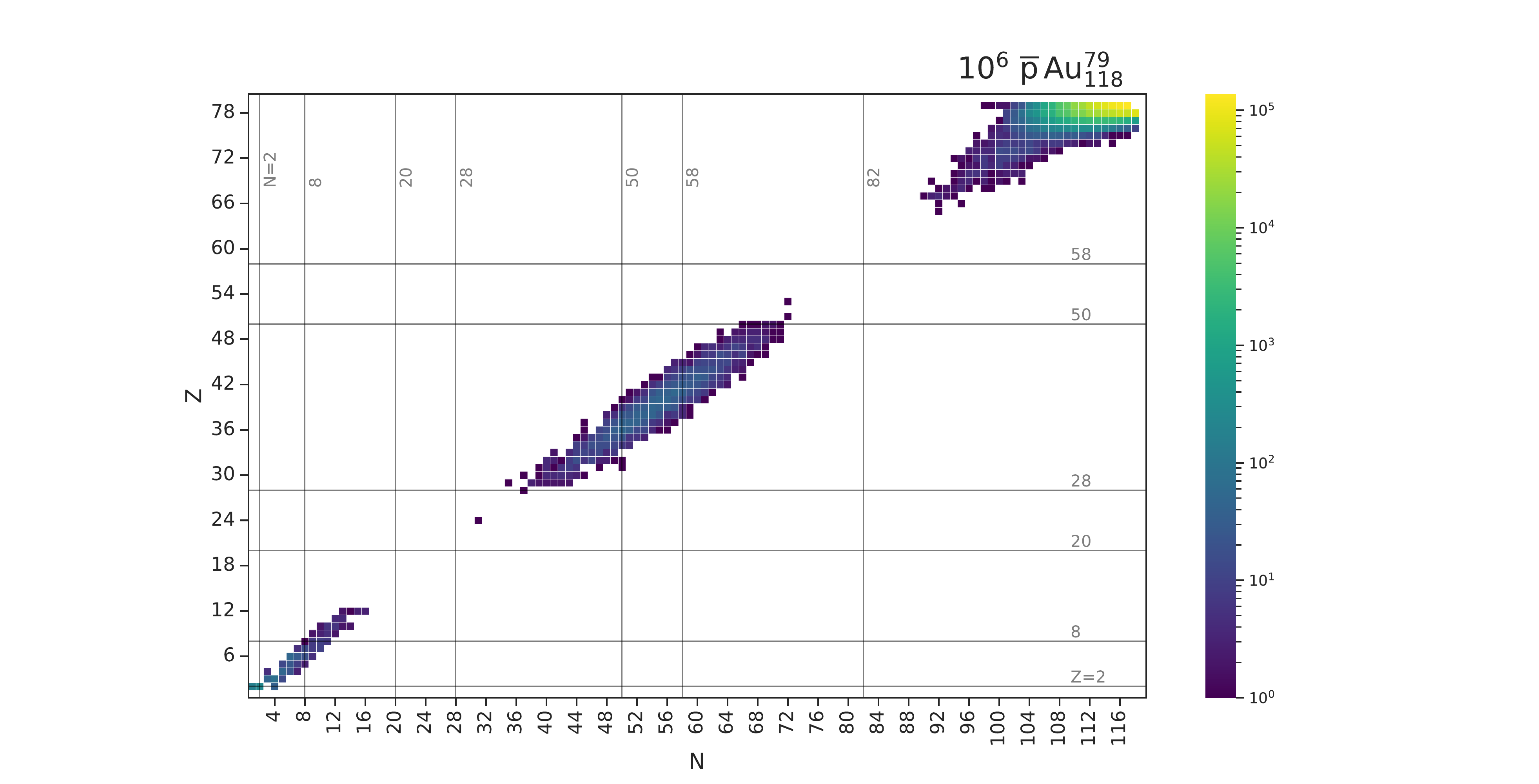}
\caption{Production rates for nuclear fragments produced in the process of annihilation of captured antiprotons by $^{197}$Au.}\label{fig:ZvsN}
\end{figure}

The simulation is run with a number of isotopically pure parent elements, ranging from fluorine to plutonium (see Table \ref{tab:rates_fragments}). 
Fig.~\ref{fig:ZvsN} shows the distribution of produced nuclear fragments regardless of their kinetic energy for $10^6$ annihilation of antiprotons on $^{197}$Au. 
Three regions can be clearly identified; the first one is in the vicinity of N and Z of the initial nucleus. 
This is the most likely outcome, especially for N-1, N-2 and Z-1 cases. 
This region is complemented by low atomic number fragments in the left region. 
The third region shows that there is a chance of fragmenting the original nuclei into two or more fragments with approximately half the atomic number of the original nucleus.  
For each nucleus produced in the simulation, the assumption is that the parent nucleus, and thus the produced fragment, is fully stripped and has a total charge Q of Z. 
For heavier nuclei, this assumption is only partly valid, as for nuclei heavier than Kr, the most deeply bound (K-shell and L-shell) electrons may not be Auger ejected in the atomic cascade (although they may well be ejected in the annihilation process); nevertheless, this only changes the charge, and thus the charge-weighted energy, of the resulting nuclear fragment by a less than 10 percent.

\begin{figure}
\centering
\includegraphics[width=0.47\textwidth, trim={2.0cm 0 5cm 0}]{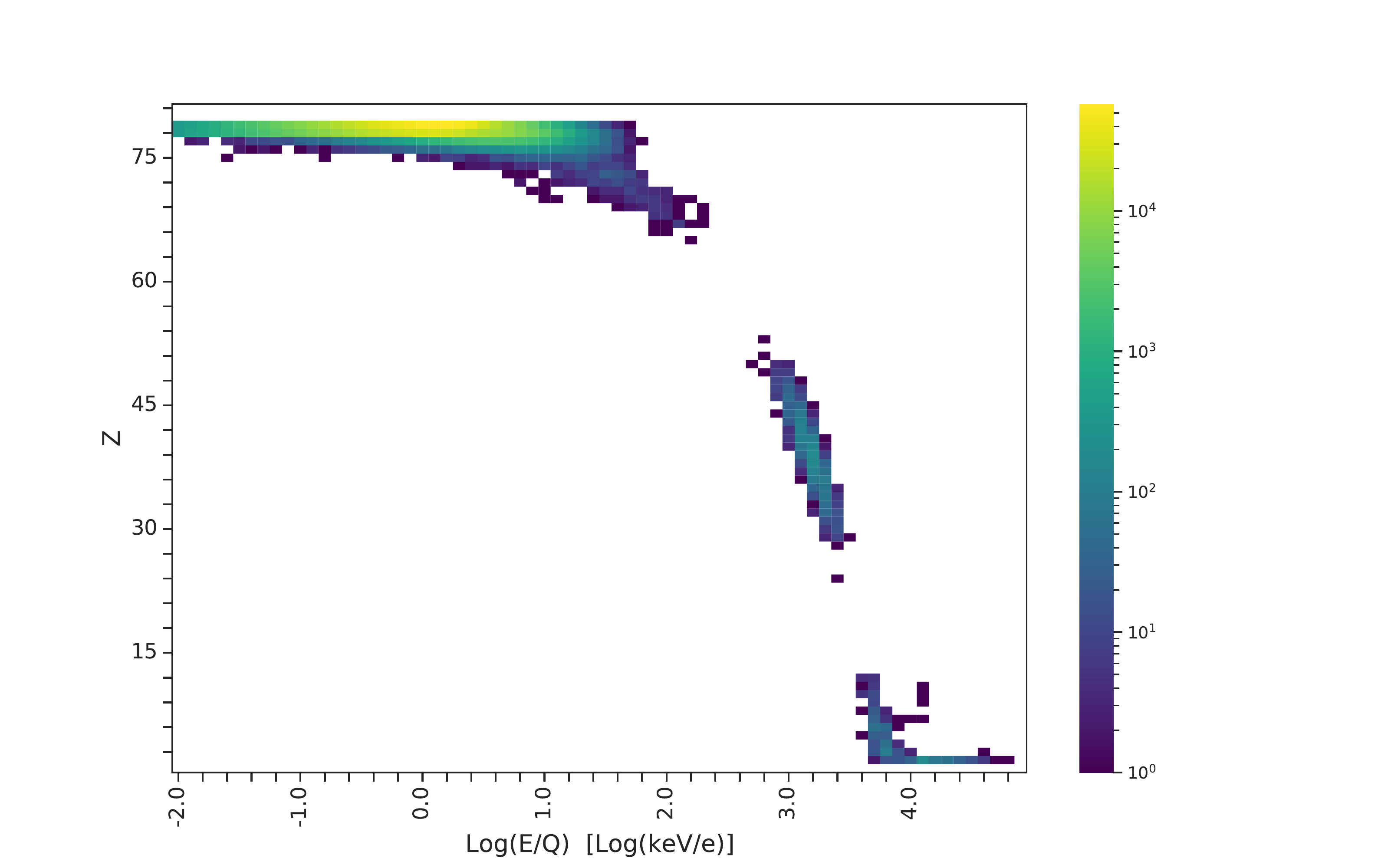}
\caption{Left: Charge-weighted energy distribution (in keV/$e$) for different species produced from 10$^6$ $\overline{\rm{p}}^{197}$Au. (Same events as in Figure~\ref{fig:ZvsN})}\label{fig:Au:E_Q}
\end{figure}

\begin{figure*}
\centering
\includegraphics[width=0.7\textwidth,trim={200pt 0 200pt 0}]{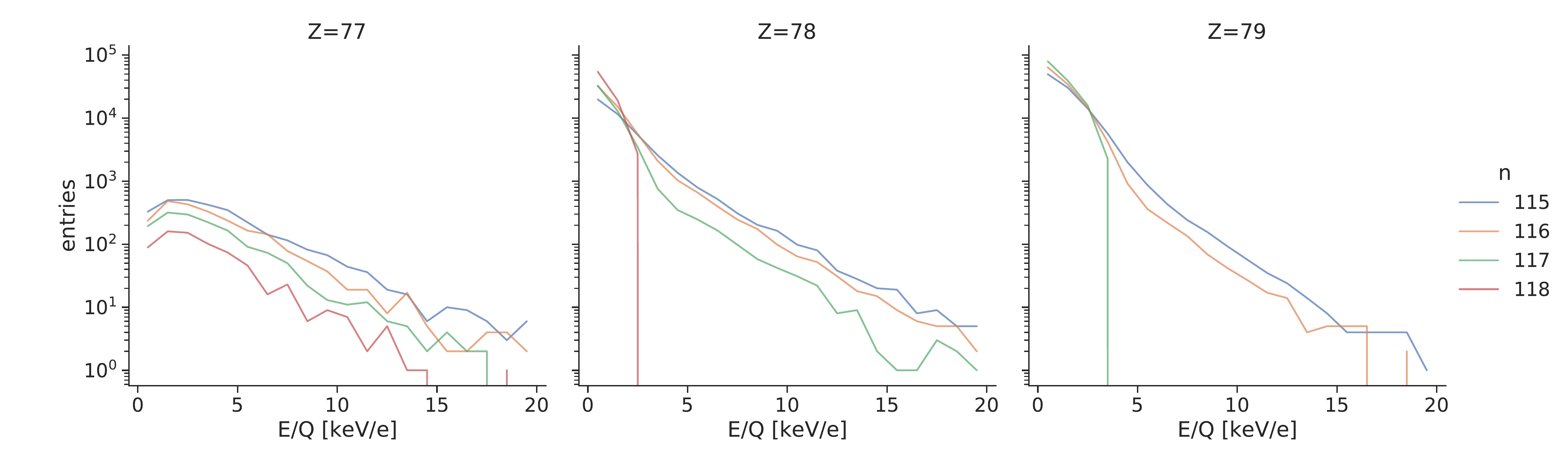}
\caption{Charge-weighted energy distribution (in keV/e) of several isotopes of Au (Z=79), Pt (Z=78), and Ir (Z=77) produced from 10$^6$ $\overline{\rm{p}}^{197}$Au.}\label{fig:E_Q}
\end{figure*}

The energy distribution of fragments produced in capture and annihilation at rest of antiprotons on gold nuclei as a function of Z can be found in Fig.~\ref{fig:Au:E_Q}. 
The three regions have a very characteristic behaviour attending to their energy normalized by the fragment's charge. 
The high-Z nuclei have energies below 100 keV/$e$, the intermediate mass have energies of the order of 1 MeV/$e$ and light fragments have energies in the range of 10-100 MeV/$e$.
Fig.~\ref{fig:E_Q} further differentiates the value of energy over charge E/Q for several isotopes of Au, Pt ad Ir produced from a $^{197}$Au target. 
The steep energy distributions show that most of the fragments have low kinetic energies and the smaller the number of evaporated nucleons is, the lower the fragment's energy.  

The scaling by the charge of the fragment is done to evaluate the possibility of trapping them immediately after their production in a Penning trap. 
As the relevant quantity is the E/Q ratio, this allows one to set an energy range in which trapping of the fragments is feasible. 
The selected value is a potential of 10 kV, which is technologically achievable in experiments~\cite{Krasnicky:2012bkl}. 

\begin{figure*}
\centering
\includegraphics[width=0.85\textwidth,trim={2cm 0 2cm 0}]{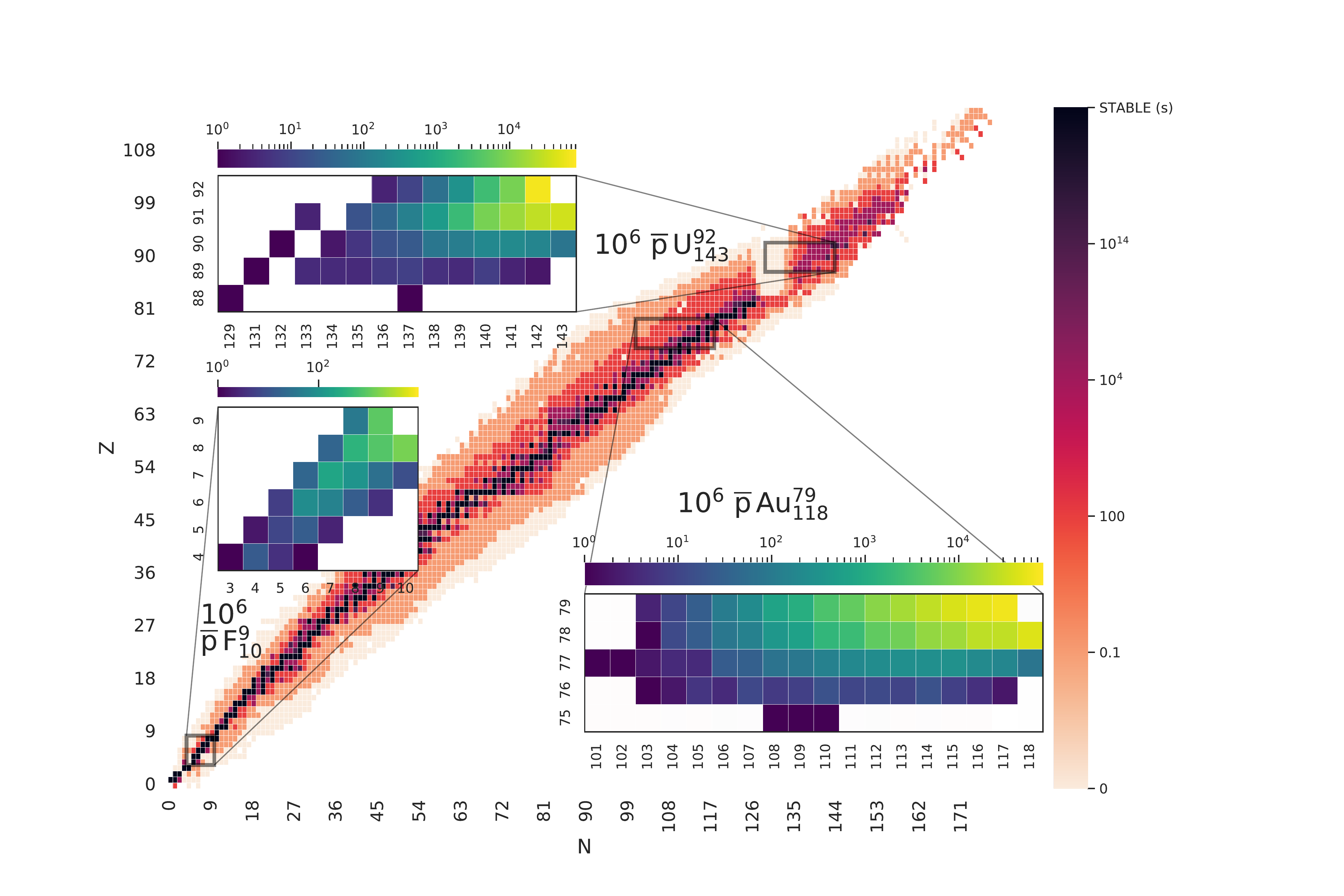}
\vspace{-10pt}
\caption{Production rates for trappable stripped fragments with energies below 10 keV/e produced from elements such as $^{235}$U, $^{197}$Au, and $^{19}$F. The technique favours evaporation of surface neutrons over protons. }\label{fig:frag_au_u_f}
\end{figure*}

The limit of 10 keV/Z is applied to select the produced fragments in order to evaluate the trappable fraction. This analysis is shown in Fig.~\ref{fig:frag_au_u_f} for Au, U and F initial atoms. 
For the heavy systems such as Au and U, more than 90 \% (see Table \ref{tab:rates_fragments}) of the fragments have energies below the threshold and are thus trappable. 
Elements from Z to Z-4 and N to N-17 can be produced. The closer to the initial number the more abundant they are. 
In case of the lighter F, only few \% could be captured in a potential of 10 kV, as in general, these fragments have higher energies. 

\begin{table}[]
    \centering
    \caption{Synthesis rate in \% of several trappable (E/Q$<$10~{\si{\kilo\volt}}) highly charged isotopes for different initial atomic systems. While a large fraction of HCI's stemming from stable parent atoms is trappable, in the case of radioactive parent atoms, spontaneous nuclear spallation of the remnant produced by $\overline{\rm{p}}$ annihilation reduces this fraction noticeably.}
    \begin{ruledtabular}
    \begin{tabular}{c|c|c|c|c}
        Initial & tra* & N-1 & Z-1 & Others \\
         \hline
        $^{19}$F & 2 & 0.5 & 1 & -- \\
        $^{59}$Co & 40 & 7.5 & 7.0 & $<$0.5($^{50--52}$Co) \\
        $^{165}$Ho & 96 & 13.0 & 7.5 & 12.0($^{163}$Ho) \\
        $^{181}$Ta & 97 & 12.5 & 7.0 & 12.0($^{179}$Ta), 10.0($^{178}$Ta)\\
        $^{184}$W & 98 & 13.0 & 6.0 & 12.0($^{182}$W), 9.0($^{181}$W)\\
        $^{188}$Os & 98 & 13.0 & 6.5 & 12.5($^{186}$Os) \\
        $^{197}$Au & 98 & 13.5 & 7.5 & -- \\
        $^{210}$Po & 98 & 15.0 & 6.5 & -- \\
        $^{222}$Rn & 97 & 10.0 & 4.5 & 9.0($^{218}$Rn), 6.0($^{217}$Rn), 6.0($^{216}$Rn)\\
        $^{226}$Ra & 88 & 11.5 & 5.0 & 4.5--2.0($^{220-218}$Ra), 2.5--1.5($^{219-217}$Fr)\\
        $^{232}$Th & 56 & 11.5 & 5.0 & 4.0($^{229}$Th) \\
        $^{235}$U & 28 & 11.0 & 5.0 & $<$0.5($^{222-225}$U)\\
        $^{238}$Pu & 17 & 9.5 & 5.0 & -- \\
        $^{244}$Pu & 28 & 10.0 & 4.5 & --  
    \end{tabular}
    \end{ruledtabular}
    \label{tab:rates_fragments}
\end{table}

\section{Discussion} \label{Discussion}

These simulations show the feasibility of a new scheme for the  formation and in-situ trapping of radioisotopic HCI's through the laser-stimulated interaction of co-trapped antiprotons with negative ions of stable or long-lived elements in a single or in a nested Penning trap. 
This method provides access to a range of trapped, completely (or almost completely) stripped radionuclei as well as a precise time tagging of each synthesis event. 
While their energies upon formation are still in the range of keV, they can be sympathetically cooled to O(K) temperatures (by incorporating into the same traps a plasma of positrons~\cite{Andresen2011}) or to {\si{\micro\kelvin}} temperatures (by laser cooling of co-trapped positive ion species~\cite{Larson1986, Schmoeger2015}).

Our synthesis scheme is general and adapts to many atomic and molecular precursors, including some that are currently difficult to obtain from e.g. spallation sources, such as W or Ta. 
Additionally, the study of isotopes of radioactive elements (U, Po, Pu) can be simplified, given the very small numbers of required parent atomic ions.
The method can produce HCIs consisting of a mix of around a dozen isotopes of two to four elements, which will be amenable to further manipulations. 
Assuming initially 10$^6$ antiprotons co-trapped with 10$^6$ isotopically pure anions that are neutralized and brought to a Rydberg state, around 1000 trappable HCI's are produced~\cite{Gerber}.  
Eliminating undesirable isotopes from among the trapped mix can be carried out by superimposing appropriate RF on the Penning trap's trapping potentials~\cite{Dilling:2018mqa}.
Determination of precise trapping efficiencies lies beyond the focus of this paper, given uncertainties inherent in the simulations of the underlying physics processes: GEANT4 does not e.g. address possible ejection of deeply bound electrons (that are not Auger ejected during the atomic cascade of heavy antiprotonic atoms) through the strong local electric field fluctuations during annihilation. 
Similarly, different simulations codes (e.g. FLUKA~\cite{FLUKA}) could result in different isotopic production rates through different implementations of the intranuclear cascade. 
Nonetheless, these uncertainties will not reduce the fraction of trappable radioisotopic HCI's significantly.


A number of implications of this formation scheme can be considered, such as the 
synthesis, trapping and cooling of particular radioisotopes that are currently of great interest, but are difficult to produce, such as $^{229}$Th, which is a prime candidate for a nuclear clock~\cite{Jeet:2015jna,vonderWense:2016fto}. 
As shown in Tab.~\ref{tab:rates_fragments}, around 5~\% of antiprotonic-$^{232}$Th atoms result in the trappable radio-isotopic HCI of $^{229}$Th, potentially at meV energies (if sympathetic laser cooling is implemented); this trapped and cooled $^{229}$Th can either be studied in situ, transported towards precision traps or be extracted and accumulated in view of transport~\cite{Tseng:1993iii} to off-laboratory sites. 
Other equally relevant radio-isotopes are $^{163}$Ho (that could be produced and trapped via the same formation scheme from $^{165}$Ho), and to a lesser extent, given the small amounts that could be expected to be produced, $^{187}$Re (from $^{188}$Os), both of which are at the focus of experimental attempts to measure the neutrino mass.

Similarly, a range of proton-rich short-lived (lifetimes of 10 ms or less) radioisotopes become accessible since the formation time is known with O(100 ns) accuracy~\cite{Amsler2021}, and trap manipulations (e.g. to separate out unwanted isotopes) can take place on timescales of few microseconds. Specifically, this method potentially provides access to small numbers of trappable $^{215}$At,$^{216}$At,$^{217}$At as well as $^{216}$Rn, $^{217}$Rn and $^{218}$Rn (when starting from the parent $^{222}$Rn, with a lifetime of 2 days); of $^{217}$Fr,$^{218}$Fr,$^{219}$Fr or $^{217}$Ra, $^{218}$Ra, $^{219}$Ra and $^{220}$Ra (when starting from the parent $^{226}$Ra, with a lifetime of 1600 years); or of $^{221}$Pa, $^{222}$Pa, $^{223}$Pa and $^{222}$U, $^{223}$U, $^{224}$U and $^{225}$U (when starting from the parent $^{235}$U with a lifetime of 10$^8$ years). But also production and trapping of short-lived isotopes of much lower-mass elements, such as $^{50}$Co,$^{51}$Co or $^{52}$Co should be feasible.

Accessing neutron-rich short-lived isotopes is significantly more difficult, as the simulations indicate that processes involving emission of multiple neutrons are much more likely than those involving more than one (or maximally two) emitted protons (and those generally accompanied by multiple neutrons). 
Nevertheless, since the $\Delta$Z=$\Delta$N=1 process has a rather large likelihood, a two step process can be envisaged: first formation of trapped $^{\rm{N}-1}_{\rm{Z}-1}$A$^{(\rm{Z}-1)+}$ from antiprotonic-$^{\rm{N}}_{\rm{Z}}$A, followed in a second step by a second injection of further antiprotons into the trapped (and purified) $^{\rm{N}-1}_{\rm{Z}-1}$A$^{(\rm{Z}-1)+}$ HCI's. 
While the formation of antiprotonic atoms in this configuration requires three-body interactions, 
and thus does not allow pulsed formation, a small amount of the trapped population would result in a further $\Delta$Z=$\Delta$N=1 reaction, producing $^{\rm{N}-2}_{\rm{Z}-2}$A$^{(\rm{Z}-2)+}$. 

More generally, the availability of positrons or positive ions optically cooled to {\si{\micro\kelvin}} (for sympathetic cooling~\cite{cooling_science_2015}) and recent experience in pulse-forming and laser-exciting positronium~\cite{AEgIS:2021rqk}, allow one to dress any trapped (and fully or almost fully stripped) HCI with an electron via formation and charge-exchange interaction of pulse-formed positronium with the trapped HCI. 
If the positronium is in a Rydberg state, so will be the electron around the HCI, thus allowing to probe QED in the strong field regime as well as to search for novel electro-weak like interactions, through precision spectroscopy between Rydberg states in a hydrogen-like configuration in which two-body calculations are sufficient, and in which nuclear form factors are negligible~\cite{Kozlov:2018mbp}. This is of particular interest for EDM searches in non-symmetric nuclei like $^{225}$Ra, $^{229}$Th or $^{229}$Pa~\cite{228Th} (all of which would be accessible from longer-lived parent nuclei in our scheme).
Alternatively, as proposed in~\cite{DOSER2022103964}, nearby pulsed formation of Rydberg protonium (instead of positronium), or of any other antiprotonic atom in a Rydberg state~\cite{Gerber} would, via the reaction $^{\rm{N}-1}_{\rm{Z}-1}$A$^{(\rm{Z}-1)+} + \rm{Pn}^* \rightarrow \overline{\rm{p}}-^{\rm{N}-1}_{\rm{Z}-1}$A$^{(\rm{Z}-2)+}{}^* + p$, allow pulsed formation of hydrogen-like antiprotonic Ryberg ions of short-lived, fully stripped, HCI's, of similar interest for fundamental physics~\cite{Kozlov:2018mbp}.

\section{Conclusions} \label{Conclusions}

We have proposed and investigated a scheme based on pulsed production of antiatomic atoms that offers a novel and potentially interesting route for the production of a range of trappable radio-isotopes in form of fully or almost fully stripped HCI's that are of potential interest to fundamental studies (tests of QED, searches for Z'), formation of trapped exotic hollow HCI's with a single electron or antiproton in a Rydberg state, or production of trapped, short-lifetime radioisotopes amenable to investigation on time scales of {\si{\micro\second}} after their formation.

While previous data  on antiproton-induced production of long-lived radioisotopes, as well as the simulations of the present investigation, indicate that this production scheme should be both effective and versatile, experimental verification with co-trapped ions and antiprotons is currently only in the planning stage and will thus only become possible in a few years.


\acknowledgments 

The work was funded by Warsaw University of Technology within the Excellence Initiative: Research University (IDUB) programme and the IDUB-POB-FWEiTE-1  project grant as well as a part of a project co-financed by the Minister of Education and Science on the basis of an agreement No. 2022/WK/06. 

This work has received funding from the European Union’s Horizon 2020 research and innovation programme under the Marie Skłodowska-Curie grant agreement No 801110 and the Austrian Federal Ministry of Education, Science and Research (BMBWF). It reflects only the authors' view; the EU agency is not responsible for any use that may be made of the information it contains.


\renewcommand{\thefigure}{A.\arabic{figure}}
\setcounter{figure}{0}
\renewcommand{\theequation}{A.\arabic{equation}}
\setcounter{equation}{0}
\section*{Appendix A: Antiprotonic atoms from an anion beam}
\begin{figure}
\centering
\includegraphics[width=0.47\textwidth]{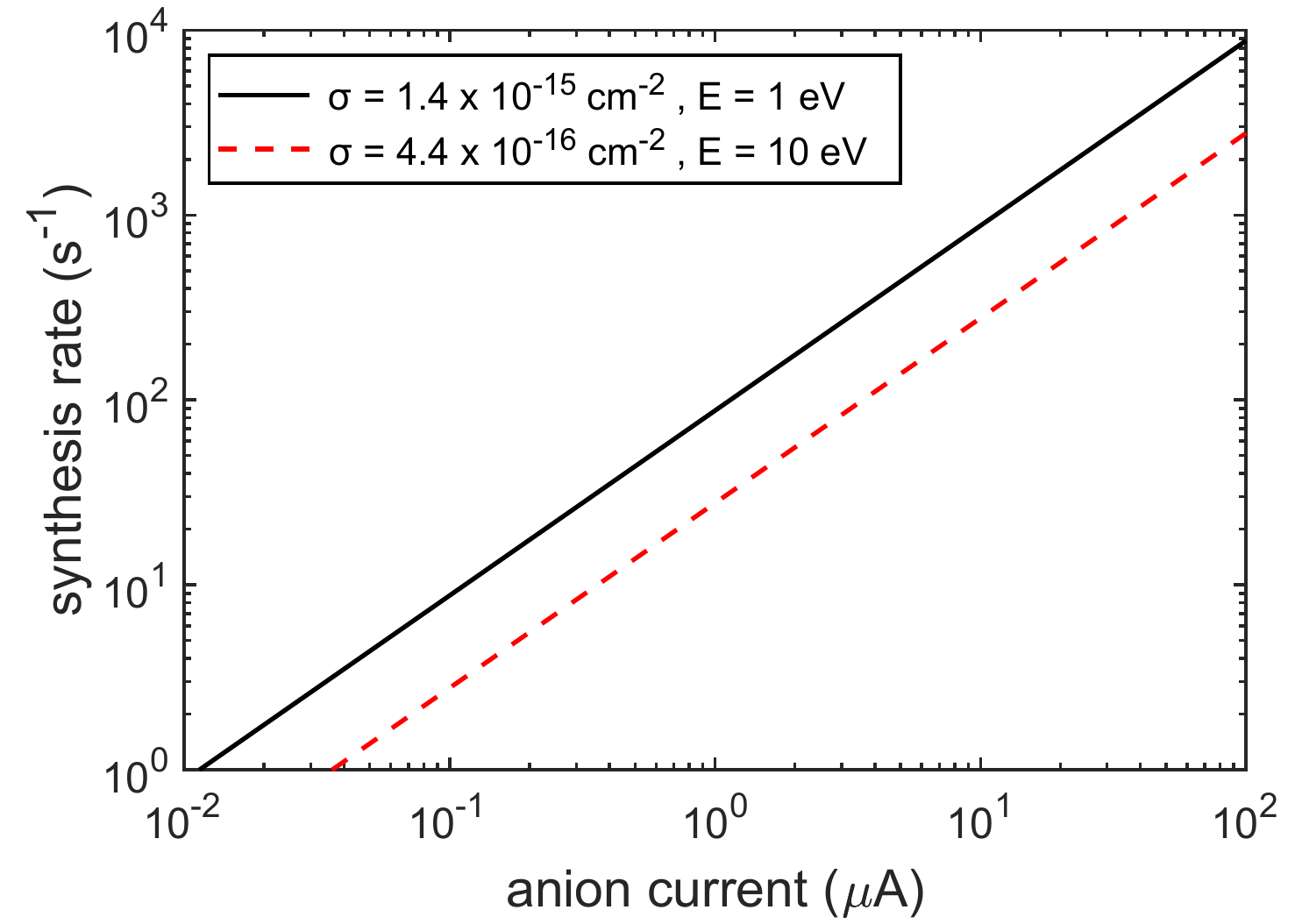}
\caption{Formation rate of protonium predicted from the interaction of trapped antiprotons with a beam of $\textrm{H}^-$ ions. The rate is calculated assuming the anion beam and the antiproton interacting on the total area of $S=0.785$~cm$^2$. The legend labels indicate the collision energy $E$ in the center-of-mass frame with the corresponding interaction cross-section $\sigma$.}
\label{fig:formationrate}
\end{figure}
In this appendix, we estimate the production rate of antiprotonic atoms obtainable by streaming a beam of anions on trapped antiprotons. 
We will assume $N=8 \times 10^5$ antiprotons trapped in a Penning trap with cylindrical electrodes within a radius of $a_r = 5$~mm from the axis as reported in Ref.~\cite{Amsler2021}. 
Anions can be produced as a beam from a Cs sputter source and streamed on-axis and at energies of several kV in experimental conditions similar to the ones described in Ref.~\cite{Cerchiari2018} through the Penning trap. 
The electrodes defining the antiproton trapping region are floated to match the energy of the incoming beam, allowing to adjust the collision energy of the antiprotons and anion between 1~eV and a few tens of eV. 
The deceleration happens inside the magnetic field of the trap, which prevents the anion beam from diverging even at the low energies of a few eV. 
Thus, the cross-section of the antiproton cloud and the anion beam can be approximated by $S = \pi a_r^2 = 0.785$~cm$^2$. 
The collision energies should be between 1 eV and 10 eV. 
Collisions happening at lower energies are unlikely to detach the extra-electron from the negative ion, in which case the Coulomb repulsion separates the antiprotons preventing the reaction with the matter atoms~\cite{Gerber}. 
Collisions at higher energy suffer from a dramatic reduction of the interaction cross section and are also unlikely to form antiprotonic atoms~\cite{Morgan:1970yz}.
An estimate for the interaction cross-section at a few eV for the formation of antiprotonic atoms can be found in previous experiments aiming to measure the rest gas pressure in the vacuum chamber by measuring the rate of annihilation of trapped antiprotons~\cite{Sellner_2017_pbarlifetime, Feithesis}. Figure~\ref{fig:formationrate} presents the expected annihilation rate for different collision energies of $\textrm{H}^-$ with antiprotons as an exemplary anions. 
We see that with an anion current of 10~\si{\micro\ampere}, which can be produced by using a Cs sputter source, the rate of protonium synthesis could be in the order of $\sim10^2$~s$^{-1}$. 
This rate could be higher for other elements than H because it is predicted that heavier elements should have higher interaction cross-section~\cite{Cohen}.

\bibliography{apssamp} 

\end{document}